\documentclass[draftcls,onecolumn]{ieeeconf}

\IEEEoverridecommandlockouts                              

\usepackage{graphicx} 
\usepackage{amsmath} 
\usepackage{amssymb}  
\usepackage{listings}
\usepackage{color}
\usepackage{algorithm}
\usepackage{pgf}
\usepackage{color}
\usepackage[latin1]{inputenc}
\usepackage[noend]{algpseudocode}
\usepackage{subcaption}
\usepackage{url}

\graphicspath{{Images/}}
\newtheorem{dfn}{Definition}
\newtheorem{lemma}{Lemma}
\newtheorem{problem}{Problem}
\newtheorem{rmk}{Remark}

\newcommand{\adj}{\textnormal{Adj}}
\newcommand{\Adj}{\adj}
\newcommand{\N}{\mathbb{N}}

\newcommand \La{\mathcal{L}}

\title{\LARGE \bf Towards Differential Privacy for Symbolic Systems}

\author{Austin Jones, Kevin Leahy, Matthew Hale%
\thanks{Austin Jones and Kevin Leahy are with the Massachusetts Institute of Technology Lincoln Laboratory, Lexington, MA USA. \newline
\indent Matthew Hale is with the Department of  Mechanical and Aerospace Engineering at the University of Florida, Gainesville, FL USA. \newline
\indent DISTRIBUTION STATEMENT A. Approved for public release. Distribution is unlimited.\newline
\indent This material is based upon work supported by the United States Air Force under Air Force Contract No. FA8702-15-D-0001. Any opinions, findings, conclusions or recommendations expressed in this material are those of the author(s) and do not necessarily reflect the views of the United States Air Force.
}%
}

\begin{document}

\maketitle

\begin{abstract}
In this paper, we develop a privacy implementation for symbolic control
systems. Such systems 
generate sequences of non-numerical data, and these sequences
can be represented by words or strings over a finite alphabet.
This work uses the framework of differential privacy, which is a
statistical notion of privacy that makes it unlikely that privatized data will reveal
anything meaningful about underlying sensitive data. To bring differential
privacy to symbolic control systems, we develop an exponential mechanism
that approximates a sensitive word using a randomly chosen
word that is likely to be near it.  
The notion
of ``near'' is given by the Levenshtein distance, which 
counts the number of operations required to change one string into another. 
We then develop a Levenshtein automaton
implementation of our exponential mechanism that efficiently generates
privatized output words. This automaton has letters as its states, and
this work develops transition probabilities among these states that give overall output
words obeying the distribution required by the exponential mechanism. 
Numerical results are provided to 
demonstrate this technique for both strings
of English words and runs of a deterministic transition system, demonstrating
in both cases that privacy can be provided in this setting while maintaining
a reasonable degree of accuracy. 
\end{abstract}
\section{Introduction}
Control systems appear in a wide range of applications and are used
in a wide range of problem formulations. As control applications have become increasingly 
reliant upon user data, there has arisen interest in protecting individuals' privacy
in some applications, e.g., in smart power grids~\cite{simmhan11,siddiqui12}. 
In response, there has been some work on privacy in control, 
and sensitive
user data have been made private in
multi-agent control systems~\cite{cortes16,wang17}, 
convex optimization~\cite{nozari18,hale17,han17},
linear-quadratic control~\cite{hale18}, 
and a range of filtering and estimation 
problems~\cite{leny14,leny18}.
All of these problems protect 
sensitive numerical data by adding carefully calibrated noise to such data before
they are shared.

However, methods based on additive noise do not readily extend to non-numerical data, nor to sequences of them. 
Symbolic control systems generate sequences of non-numerical
data, which are analogous to trajectories for ordinary control
systems, and these are typically represented as words or strings over a finite
alphabet. 
A symbolic trajectory can represent, for example, 
a switching sequence prescribing modes to switch
between in a hybrid control system~\cite{tabuada09}, or
a sequence of finite subsets of state space to occupy, such as in a path
planning problem~\cite{belta07}. 

A symbolic trajectory
can therefore reveal one's actions or positions over time, and this setting incurs
privacy concerns similar to those with trajectories of numerical
data. Simultaneously, agents may need to share these trajectories with other
agents to jointly coordinate their activities, though a network may contain
untrusted agents, or communications may be subject to eavesdropping. 
Thus there is a need to share symbolic trajectories
in a way that preserves their accuracy while providing strong,
provable privacy guarantees to users. Because
existing approaches do not readily extend to the symbolic setting,
fundamentally new approaches are required to 
ensure that sensitive symbolic data of this kind can safely be shared. 

Accordingly, in this paper we develop a general-purpose method for providing privacy
to sensitive words generated by symbolic control systems. To do so, we adopt the
framework of differential privacy. 
Differential
privacy is a statistical notion of privacy that makes it unlikely for an eavesdropper
or adversary to learn anything meaningful about sensitive data from its differentially
private form~\cite{dwork14}. 
Its key features include immunity to
post-processing, in
that transformations of privatized data to not weaken privacy
guarantees, and robustness to side
information, in that learning additional information about
data-producing
entities does not weaken differential privacy by much. 

Differential privacy originates in the database literature
in computer science, and it has been applied to protect individual database entries
in response to queries of the database as a whole~\cite{dwork06}. 
It was later extended to trajectory-valued data and
applied in the control setting in~\cite{leny14}, and has seen applications
in both its database and trajectory forms in a range of control 
settings~\cite{cortes16,wang17,hale18,leny14}. 
Differential privacy is most commonly implemented using the Laplace and Gaussian
Mechanisms, which add Laplacian and Gaussian noise, respectively, to
sensitive data before sharing them. 

Differential privacy has also been applied to non-numerical data using the
exponential mechanism, which randomly generates responses to non-numerical
queries based on how well those responses approximate 
the non-private response~\cite{mcsherry07}. 
The exponential mechanism has been applied, for example, to
data aggregation~\cite{eigner14} and data release~\cite{hardt12} problems,
as well as pricing and auction problems~\cite{dwork14,mcsherry07}. 
To bring privacy to the symbolic control setting,
we will develop an exponential mechanism for words over a finite
alphabet. 

Doing so first requires defining differential privacy
in a manner that captures the privacy needs of symbolic control
systems,
and the first contribution of this paper is formally defining
differential privacy for this setting. Next, actually implementing
differential privacy 
requires defining ``quality'' in a meaningful way.
The notion of ``quality'' we use is 
based on the Levenshtein distance from a word.
The Levenshtein distance counts how many insertions, substitutions,
and deletions are required to change one word into another.
Given a sensitive word (representing a sensitive
symbolic trajectory), our differential privacy implementation
therefore outputs nearby words (in the Levenshtein sense) with
high probability and, conversely, outputs distant words
with low probability. 

The second contribution of this paper then comes from the exponential
mechanism itself, and, for a given sensitive word, we provide
the distribution over possible output words required to implement
differential privacy. A na\"{i}ve approach to generating samples from
this distribution would iterate over all possible output words, 
compute their quality scores, and then select one as a private output.
However, actually executing these operations can be very
computationally demanding, and this na\"{i}ve implementation would
require computing all pair-wise Levenshtein distances among all possible
output words, which can incur substantial computational expense.  

Instead, to preclude the need for large scale computation,
we specify a more efficient means of generating output words, and
this implementation constitutes this paper's third contribution.
In particular, we construct a Levenshtein automaton that 
generates output words one letter at a time in a manner
that obeys the probability distribution required by differential
privacy. The state of the Levenshtein automaton is defined to be
the letter most recently added to the output word, and transition
probabilities are constructed between letters to determine which letter
should be added to the output word next. Levenshtein automata
can be constructed efficiently, and our implementation provides
a substantial computational improvement over the na\"{i}ve approach.
In this preliminary study, we consider a restricted form of the
Levenshtein distance, namely, we allow substitutions but not
deletions or insertions, and we defer the use of the full Levenshtein
distance to a future publication.  
These modifications require only removing
certain transitions from the automata that we construct, 
and this can likewise be done efficiently, as will be shown.
This implementation will be demonstrated on both English words 
and a deterministic transition system to show its applicability
in symbolic control systems and beyond.

We note that a Levenshtein automaton can be represented as a graph, 
with each letter a node in the graph 
and the transition probabilities acting as edge weights. 
Differential privacy has been applied to graphs in various ways,
including to protect topological characteristics~\cite{hay09,kasiviswanathan13} 
and the edge weights within
a graph~\cite{pinot18}. This paper differs from those works because 
we use graphs merely for the implementation of
the exponential mechanism over words, and we are not applying
privacy to any graph structure. 

The rest of the paper is organized as follows. Section~\ref{sec:prelim}
presents the relevant background on Levenshtein automata
and differential privacy. Then, Section~\ref{sec:problem} formally states
the differential privacy problem that is the focus of the paper. 
Section~\ref{sec:mech} next defines the exponential mechanism for
words that we use, and Section~\ref{wordGen} uses Levenshtein
automata to provide an efficient means of generating samples from
this distribution. These results are demonstrated for transition systems in
Section~\ref{runTS}, where we apply our theoretical results
to the problem of generating private runs of such systems. 
Section~\ref{sec:experiments} then provides numerical results for
both strings in a general setting and the transition system setting.
Finally, Section~\ref{sec:conclusion} provides concluding remarks
and directions for future research. 


\section{Preliminaries} \label{sec:prelim}
In this section, we define our 
notation and establish some mathematical preliminaries for the 
developments below.
\subsection{Languages}

An \emph{alphabet} is a collection of symbols $\Sigma$. 
A
word over $\Sigma$ is a concatenation of symbols $w = \sigma_0\sigma_1 \ldots$
such that $\sigma_i \in \Sigma$ for all $i$. We use the notation $\Sigma^*$ to denote to the
set of all finite words over $\Sigma$. Any subset $L \subseteq
\Sigma^*$ (equivalently $L \in 2^{\Sigma^*}$) is called
a \emph{finite language}.


 \subsection{Finite State Automata}

\begin{dfn}
A (nondeterministic) \emph{finite state automaton} (NFA) is a tuple $A=\left(Q,\Sigma,q_0,\delta,F\right)$, where $Q$ is a set of states, $\Sigma$ is an input alphabet, $q_0\in Q$ is the initial state, $\delta\subseteq Q\times\Sigma\times Q$ is the 
transition relation between states, 
and $F\subseteq Q$ is the set of accepting states. 
\end{dfn}
\begin{dfn}
A \emph{run} on an NFA $A=\left(Q,\Sigma,q_0,\delta,F\right)$ induced by word $w = \sigma^0\sigma^1\ldots \in \Sigma^*$ is a word $q=q^0q^1 \ldots \in Q^*$ such that $q^0 = q_0$ and $(q^i,w^i,q^{i+1}) \in \delta$.  Automaton $A$ \emph{accepts} a word $w$ if the final state of the induced run is an accepting state $q_f \in F$.  We call the set of all words accepted by the automaton its \emph{language}, denoted by $\La(A)$.
\end{dfn}
\begin{dfn}
 A \emph{deterministic finite state automaton} (DFA) is an NFA with deterministic transition function $\delta : Q\times\Sigma\rightarrow Q$. 
\end{dfn}

\subsection{Transition Systems}
\begin{dfn}
A \emph{deterministic  transition system (DTS)} is
given as a tuple $TS = (S, s_0, Act, T)$, where
\begin{itemize} 
\item $S$ is a finite state space,
\item $s_0 \in S$ is an initial state,
\item $Act$ is an input set
\item $T : S \times  Act \to S$ is a deterministic transition function
such that applying input $a_1$ in state $s_1$ will move the
system to state $T(s_1, a_1)$.
\end{itemize}
\end{dfn}

\begin{dfn}
A \emph{plan} for a DTS is a sequence of actions $\mathfrak{a} =a^0a^1a^2\ldots \in Act^*$.  A plan $\mathfrak{a}$ is in the input language of a DTS, denoted $\mathfrak{a} \in \La_i(TS)$, if it induces a \emph{run} $r(\mathfrak{a}) = s^0s^1\ldots \in S^*$ such that $s^0 = s_0$ and 
$s^{i+1} = T(s^i,a^i)$ $\forall i$. 
\end{dfn}




\subsection{Levenshtein Distances and Automata}

To compare two strings, we introduce the notion of Levenshtein distance \cite{levenshtein} and Levenshtein automata \cite{schulz}. These tools can be used to measure the difference or edit distance between two strings.

\begin{dfn}[Levenshtein Distance~\cite{levenshtein}]
The Levenshtein distance  between words $w_1,w_2$, denoted $d_L(w_1,w_2)$, 
is the minimum number of changes---insertions, substitutions, or deletions--- that can be applied to $w_1$ to convert it to $w_2$.
\end{dfn}

  For example, the Levenshtein distance between ``sample" and ``examples" is $3$, since the ``s" at the beginning of ``sample" must be substituted for an ``e" or an ``x," the remaining letter (``e" or ``x", whichever was not substituted for the ``s") must be added, and an ``s" must be added to the end of the word. We can identify whether a string is within a specific Levenshtein distance of another string using a Levenshtein automaton.

\begin{dfn}[Levenshtein Automaton~\cite{schulz}]
For a string $x$ and a distance $k\in\mathbb{N}$, the \emph{Levenshtein automaton} is an NFA $\mathcal{A}_{x,k}=\left(Q,\Sigma,q_0,\delta,F\right)$ such that $\La(\mathcal{A}_{x,k})$ is the set of all words with Levenshtein distance less than or equal to $k$ from $x$.
\end{dfn}

For a given string $x$ and a distance $k$, we can construct the corresponding Levenshtein automaton $\mathcal{A}_{x,k}$ as follows, using Algorithm~\ref{alg::levenshtein}.

\begin{algorithm}
\caption{Levenshtein Automaton Construction}\label{alg::levenshtein}
\begin{algorithmic}[1]
\Procedure{MakeLevenshtein}{$\Sigma$, $x\in\Sigma^*$, $k\in\mathbb{N}$}
\State $Q\gets \left\lbrace q_{i,e} \mid i\in\left\lbrace 0,1,\ldots,\vert x\vert\right\rbrace, e\in\left\lbrace 0,1,\ldots,k\right\rbrace\right\rbrace$
\State $F\gets \left\lbrace q_{i,e}\in Q \mid i=\vert x\vert\right\rbrace$
\State $\delta\gets\emptyset$
\State $q_0\gets q_{0,0}$
\For{$i\in\left\lbrace 0,1,\ldots,\vert x\vert\right\rbrace$}
    \For{$e \in\left\lbrace 0,1,\ldots,k\right\rbrace$}
        \State $\delta\gets \delta\cup \left(q_{i,e},x_i,q_{i+1,e}\right)$ \Comment{Correct transition}
        \If{$l<k$}
            \State $\delta\gets \delta\cup \left(q_{i,e},*,q_{i+1,e}\right)$ \Comment{Deletion}
            \State $\delta\gets \delta\cup \left(q_{i,e},\epsilon,q_{i,e+1}\right)$\Comment{Insertion}
            \State $\delta\gets \delta\cup \left(q_{i,e},*,q_{i+1,e+1}\right)$\Comment{Substitution}
        \EndIf
    \EndFor
\EndFor
\Return $\mathcal{A}_{x,k}=\left(Q,\Sigma,q_0,\delta,F\right)$
\EndProcedure
\end{algorithmic}
\end{algorithm}

\begin{figure*}[htbp]
\begin{subfigure}[b]{0.45\textwidth}
\centering
\resizebox{\linewidth}{!}{
\includegraphics{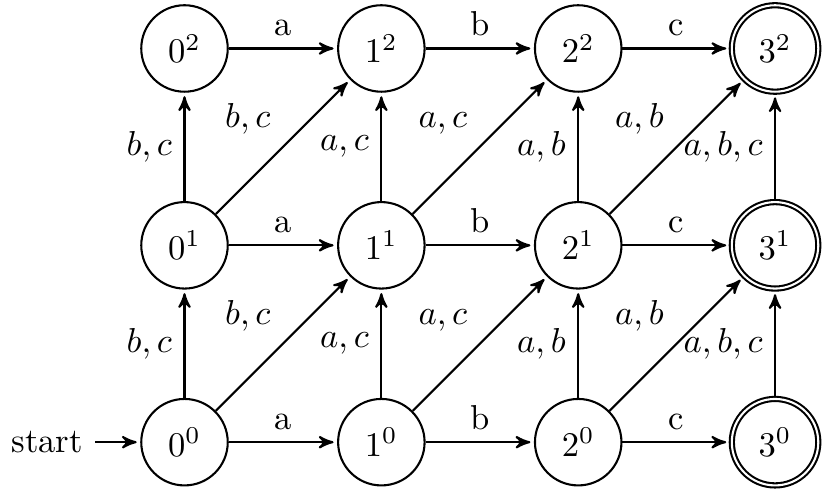}
}
\caption{}\label{fig:ex-la-a}
\end{subfigure}
\begin{subfigure}[b]{0.45\textwidth}
\centering
\resizebox{\linewidth}{!}{
\includegraphics{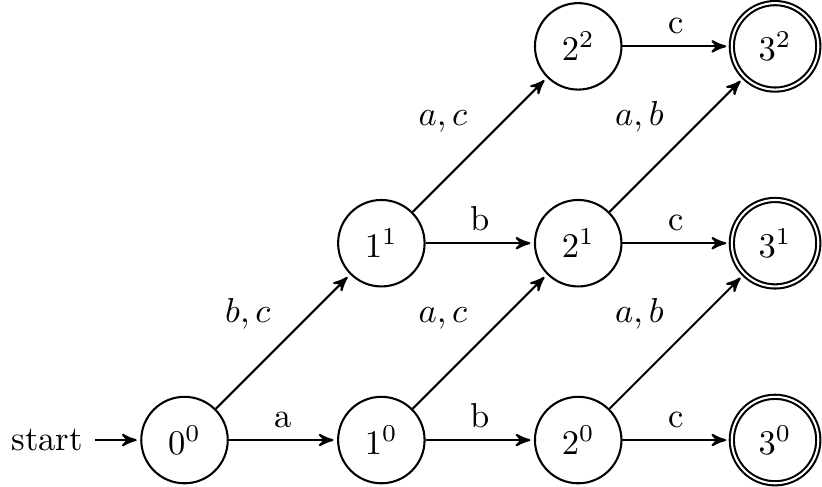}
}
\caption{}\label{fig:ex-la-b}
\end{subfigure}
\caption{(\ref{fig:ex-la-a}) Levenshtein automaton $\mathcal{A}_{x,k}$ for all words of distance less than or equal to $2$ from the word $x=abc$ from alphabet $\left\lbrace a, b, c\right\rbrace$.  Accepting states are noted with double circles. (\ref{fig:ex-la-b}) $\mathcal{A}_{k,x}^{|x|}$ for all words of length exactly $3$ and distance less than or equal to $2$ from the word $abc$ from alphabet $\left\lbrace a, b, c\right\rbrace$.}\label{fig:ex-la}
\end{figure*}

\begin{rmk}
In this paper, for simplicity, we only consider substitutions and
ignore insertions and deletions.  That is, we ignore lines 10 and 11
of Algortihm \ref{alg::levenshtein} when constructing the automaton.
We call this distance the \emph{substitution Levenshtein distance}
$d^s_L$, which is equal to the Hamming distance~\cite{hamming50}.  
Full considerations of insertions and deletions is a topic of future research.
\end{rmk}
\begin{rmk}
The automaton $\mathcal{A}_{x,k}$ generated by Algorithm~\ref{alg::levenshtein} can be pruned to a DFA that accepts only those words whose length is the same as the length of the input word, $\vert x\vert$, denoted $\mathcal{A}^{|x|}_{x,k}$.  
An example of this is shown in Figure \ref{fig:ex-la}.  The details of the DFA construction are beyond the scope of this paper and are therefore omitted. 
\end{rmk}

\subsection{Differential Privacy}
We provide here only a high-level discussion of differential privacy as background,
and further details will be provided in developing our privacy implementation below. 
The underlying goal of differential privacy is to make similar pieces
of sensitive data produce outputs with approximately equal probability distributions.
The definition of ``similar'' for sensitive data is specified by an adjacency
relation. 
Adjacency is frequently specified
in terms of a metric, e.g., the $\ell_p$-metric on a space of trajectories~\cite{leny14}
or the counting metric on the space of databases~\cite{dwork14}, and two pieces of
data are adjacent if the distance between them is bounded above by a pre-specified
constant. 
Differential
privacy then requires that adjacent sensitive data produce approximately
indistinguishable outputs. 

The notion of approximate
indistinguishability is made precise by specifying a relationship between
the probability distributions of outputs corresponding to adjacent inputs.
For adjacent sensitive data $D_1$ and $D_2$, a randomized
map $M$ provides $\epsilon$-differential privacy if
\begin{equation}
\textnormal{Pr}[M(D_1) \in \mathcal{S}] \leq e^{\epsilon} \textnormal{Pr}[M(D_2) \in \mathcal{S}]
\end{equation}
for all $\mathcal{S} \subseteq \textnormal{range}(M)$. The parameter
$\epsilon$ controls the degree of indistinguisability between the distributions of
$M(D_1)$ and $M(D_2)$, and thus the degree of privacy afforded to users.
Smaller values of $\epsilon$ provide stronger privacy guarantees, and
typical values range from $0.1$ to $\ln 3$. As noted in the introduction,
differential privacy is immune to post-processing, so that any transformation
of $M(D_1)$ or $M(D_2)$ is also $\epsilon$-differentially private, and
robust to side information, so that learning additional information about
a data-producing entity does not weaken this privacy by much. 

Given a privacy parameter
$\epsilon$, an adjacency relation, and some form of sensitive data, the principal
challenge in implementing differential privacy is finding the randomized
map $M$ that satisfies the above definition. Maps of this kind are called
\emph{mechanisms} for differential privacy, and we formally define the problem
of finding such a mechanism for words in the next section.
This problem will then be solved in Section~\ref{sec:mech}, which
provides the details of our privacy implementation.

\section{Problem Formulation} \label{sec:problem}
In this section, we provide the essential privacy definitions that
underlie this work, and then we formally state the problems
that are the focus of the remainder of the paper. 

\subsection{Word Differential Privacy}
Here, we define a novel concept of differential privacy, called \emph{word differential privacy}, that is appropriate for describing privacy for sequences of states in symbolic systems. 
\begin{dfn}[Word adjacency] \label{def:adj}
The adjacency relation between words $w_1,w_2 \in \Sigma^*$ is defined as
\begin{equation}
\Adj_{w,k} = \{(w_1,w_2) \mid d_L(w_1,w_2) \leq k \}.
\end{equation}
\end{dfn}
\begin{dfn}[Substition Word adjacency] \label{def:subadj}
The substitution adjacency relation between words $w_1,w_2 \in \Sigma^*$ is defined as
\begin{equation}
\Adj^s_{w,k} = \{(w_1,w_2) \mid d^s_L(w_1,w_2) \leq k \}.
\end{equation}
\end{dfn}
\begin{dfn}[Word Differential Privacy] \label{def:worddp}
Fix a probability space $(\Omega, \mathcal{F}, \textnormal{Pr})$. 
A mechanism
$M_w : \Sigma^* \times \Omega \to \Sigma^*$ is \emph{word $\epsilon$-differentially private} if  

\begin{equation}
    \begin{array}{c}
    \textnormal{Pr}_{\Omega}[M_w(w_1) \in L] \leq e^{\epsilon}\textnormal{Pr}_{\Omega}[M_w(w_2) \in L] \\
\forall (w_1,w_2) \in \Adj_{w,k} \forall L \in 2^{\Sigma^*}. 
    \end{array}
\end{equation}
\end{dfn}

\begin{dfn}[Substitution Word Differential Privacy] \label{def:subworddp}
Fix a probability space $(\Omega, \mathcal{F}, \textnormal{Pr})$. 
A mechanism
$M_w : \Sigma^* \times \Omega \to \Sigma^*$ is \emph{substitution word $\epsilon$-differentially private} if  

\begin{equation}
    \begin{array}{c}
    \textnormal{Pr}_{\Omega}[M_w(w_1) \in L] \leq e^{\epsilon}\textnormal{Pr}_{\Omega}[M_w(w_2) \in L]\\
\forall (w_1,w_2) \in \Adj^s_{w,k} \forall L \in 2^{\Sigma^*}. 
    \end{array}
\end{equation}
\end{dfn}

Essentially, a word differential privacy mechanism approximates
sensitive sequences of symbols with randomized versions of them.
These randomizations must have similar distributions for two
sequences that are nearby (in sense of Definitions~\ref{def:adj}
and~\ref{def:subadj}), and this is
captured by the relationships between probability distributions in
Definitions~\ref{def:worddp} and~\ref{def:subworddp}. Nearby 
symbolic trajectories are therefore made approximately
indistinguishable
to any recipient of their privatized forms, as well as any
eavesroppers who gain access to them, and these recipients are therefore
unlikely to determine the exact underlying sensitive word. 
This approximate
indistinguishability criterion is the basic idea behind differential
privacy, and it is this idea that we apply to symbolic trajectories
in this work. 
In this paper, we restrict ourselves to substitution word differential
privacy.  
The extension to word differential privacy is a topic of future research.


\subsection{Problems}
Here, we consider two problems involving substitution word differential privacy.  First, we consider the problem of synthesizing a  differentially private mechanism for an arbitrary sequence of characters from a given alphabet
\begin{problem} Fix a probability space $(\Omega, \mathcal{F}, \textnormal{Pr})$. 
Given an alphabet $\Sigma$ and a word $w$, find a mechanism $M_w: \Sigma^* \times \Omega \to \Sigma^*$ that is substitution word $\epsilon$-differentially private.
\end{problem}


Next, as a first step towards differential privacy for symbolic systems, we consider how to synthesize a mechanism for privatizing runs of a deterministic transition system.
\begin{problem}
\label{runPriv} Fix a probability space $(\Omega, \mathcal{F}, \textnormal{Pr})$. 
Given a transition system $TS$ and a run $x$, find a mechanism $M_{w,TS}: \Sigma^* \times  \Omega \to \La(TS)$  that is substitution word $\epsilon$-differentially private.
\end{problem}

A solution to Problem \ref{runPriv} would enable a designer to
privatize a desired run of a deterministic transition system  such that an agent that observes repeated executions can determine the desired trajectory only up to bounded precision. The price that must be paid for this is, of course, deviating from the desired run.  This relationship between performance and privacy is handled by tuning the single parameter $\epsilon$, and this will be shown in more detail in
Section~\ref{sec:experiments}. 

\section{The Exponential Mechanism for Words} \label{sec:mech}
In this section we define the exponential mechanism
for words over a finite alphabet. We first define
the notion of utility we use for our privacy 
implementation, and then we bound the sensitivity
of this utility. With this sensitivity bound
established, we then formally define the distribution
from which words should be drawn in order to preserve
differential privacy. Section~\ref{wordGen} below
then provides the means of efficiently generating samples
from this distribution.

\subsection{Utility  for Words in a Language}
An exponential mechanism is defined with respect to a utility function,
which quantifies the quality of each possible output. 
The choice of utility function here should therefore reflect the
quality of outputting a certain word in response
to a given sensitive input word. 
In this work, we seek to privatize
the input word by randomly outputting a word which is close
to it. Formalizing this idea, we now define the 
\emph{Levenshtein utility}, which simply captures
the idea that a private output is of higher quality
when it is closer to the input.

\begin{dfn} \emph{(Substitution Levenshtein Utility)}
Fix a constant $\alpha > 0$, an alphabet $\Sigma$, 
and a language $L \in 2^{\Sigma^*}$. 
Then, for an input word
$w_i \in L$, outputting the word $w_o \in L$ provides 
\emph{Substitution Levenshtein utility} equal to
\begin{equation}
u_{\alpha}(w_i, w_o) = \frac{1}{d^s_L(w_i, w_o) + \alpha}. 
\end{equation}
\end{dfn}

Here, the inclusion of $\alpha$ ensures that
$u_{\alpha}$ is always defined, and smaller values of
$\alpha$ will give higher values of $u_{\alpha}$ when
$w_o$ is close to $w_i$. This choice of utility has 
the benefit of decreasing rapidly as output words
disagree more with the input word, which will more strongly
bias the output of the exponential mechanism toward better
output words while maintaining privacy. 

\subsection{Sensitivity Bounds}
The next step in defining the exponential mechanism is to calculate
the sensitivity of the utility function $u_{\alpha}$. In particular,
for a fixed output word $w_o$, we must provide a bound on how
much $u_{\alpha}(\cdot, w_o)$ can differ across two adjacent
input words, and this bound will be used in defining the
distribution over possible output words below. 
Mathematically, we must bound the quantity
\begin{equation}
\Delta u_{\alpha} := \max_{v \in L} \max_{\substack{w_1, w_2 \in L \\
    (w_1, w_2) \in \adj^s_{w, k}}} |u_{\alpha}(w_1, v) - u_{\alpha}(w_2, v)|,
\end{equation}
and we have the following lemma that does so.

\begin{lemma}
Fix $\alpha > 0$ and $k \in \N$. Then the sensitivity of
$u_{\alpha}$ is bounded via
\begin{equation}
\Delta u_{\alpha} \leq \frac{k}{\alpha(k + \alpha)}.
\end{equation}
\end{lemma}
\emph{Proof:} Without loss of generality we may remove
the absolute value signs and set 
\begin{equation}
\Delta u_{\alpha} := \max_{v \in L} \max_{\substack{w_1, w_2 \in L \\
    (w_1, w_2) \in \adj^s_{w, k}}} u_{\alpha}(w_1, v) - u_{\alpha}(w_2, v),
\end{equation}
because we can 
relabel $w_1$ and $w_2$ to make the right-hand side non-negative.
Expanding the right-hand side, we find 
\begin{equation} \label{eq:l1runback}
\Delta u_{\alpha} := \max_{v \in L} \max_{\substack{w_1, w_2 \in L \\
    (w_1, w_2) \in \adj^s_{w, k}}} \frac{1}{d^s_L(w_1, v) + \alpha} - \frac{1}{d^s_L(w_2,
  v) + \alpha},
\end{equation} 
and its non-negativity requires that
$d^s_L(w_1, v) \leq d^s_L(w_2, v)$. To reduce the number of variables
in the maximization, we set
\begin{equation} \label{eq:l1cback}
d^s_L(w_2, v) = d^s_L(w_1, v) + c
\end{equation}
where $c \geq 0$. 
Using the triangle inequality we have
\begin{equation}
d^s_L(w_2, v) \leq d^s_L(w_2, w_1) + d^s_L(w_1, v) \leq d^s_L(w_1, v) + k,
\end{equation}
which follows from the adjacency of $w_1$ and $w_2$. 
Combining this with Equation~\eqref{eq:l1cback}
gives
\begin{equation}
d^s_L(w_1, v) + c \leq d^s_L(w_1, v) + k,
\end{equation}
which gives $c \in \{0, \ldots, k\}$. 

Returning to Equation~\eqref{eq:l1runback} we find that
\begin{align}
\Delta u_{\alpha} &:= \max_{v \in L} \max_{\substack{w_1 \in L \\ c
    \in \{0, \ldots, k\}}} \frac{1}{d^s_L(w_1, v) + \alpha} - \frac{1}{d^s_L(w_1,
  v) + c + \alpha} 
\end{align}
For every $v \in L$, 
maximizing over $w_1$ is easily done by setting
$w_1 = v$, which now gives
\begin{equation} \label{eq:l1almost}
\Delta u_{\alpha} := \max_{c \in \{0, \ldots, k\}} 
\frac{c}{\alpha(c + \alpha)},
\end{equation}
where we have removed the maximization over $v$ because
all dependence upon~$v$ is now eliminated.
The right-hand side in Equation~\eqref{eq:l1almost}
is maximized by maximizing $c$, which 
completes the proof. 
\hfill $\blacksquare$

Although we frequently expect this bound to be attained, 
we write it as an inequality to account
for the case that $L$ does not contain any words
exactly distance $k$ apart. 

\subsection{Distribution Over Output Words}
Given the above bound on sensitivity, the final step
needed to define the exponential mechanism is
determining the required distribution over output words.
The standard definition of the exponential mechanism~\cite{dwork14}
says that, for a given sensitive input word $x$,
a candidate word $w$ should be output with probability
$p(w; x)$ satisfying the proportionality relation
\begin{equation} \label{eq:nokx}
p(w; x) \sim \exp\left(\frac{\epsilon u_{\alpha}(x, w)}{2 \Delta u_{\alpha}}\right).
\end{equation}
For the case of word adjacency (cf. Definition~\ref{def:adj}), one 
would need to determine a proportionality constant $K_x$, and
this would require computing the distance to every possible
output word from
every possible sensitive input word~$x$. 
Computing the Levenshtein distance has 
time complexity $\mathcal{O}(n^2)$~\cite{wagner74}, though the implied constants
can be very large for longer strings and large alphabets.
Computing all possible pairwise distances can therefore
easily become intractable. 
However, for substitution word adjacency
(cf. Definition~\ref{def:subadj}),
which is the focus of this paper, explicitly computing
$K_x$ can be avoided, and this is shown in the next section. 
\section{Generating Differentially Private Words from A Fixed Alphabet}
\label{wordGen}
In the previous section, we defined an exponential mechanism for 
substitution word $\epsilon$-differential privacy.  
In this section, we propose an efficient method for generating samples $w' \sim p(\,\cdot\,;x)$. We propose to do this by synthesizing appropriate randomized policies $\mu_{\epsilon,x}:Q \times \Sigma \to [0,1]$ over the Levenshtein automaton associated with word $x$, and these policies will randomly select each letter in an output
word to implement the exponential mechanism for words. Formally,
we have the following formulation. 

\begin{problem}
Given a Levenshtein automaton $\mathcal{A}^{|x|}_{k,x}$ as constructed by Algorithm \ref{alg::levenshtein}, synthesize a policy $\mu_{\epsilon,x}$ such that  
\begin{equation}
 \prod_{e=0}^{|x|-1}\mu_{\epsilon,x}(q^e,\sigma^e) = p(\sigma^{0:k};x).
\end{equation}
\end{problem}

Note that in this approach, we are restricting ourselves to generating
privatized output words $w$ that are the same length as $x$ and are allowing
arbitrary symbols from an alphabet $\Sigma$ to be selected.  That is,
we are privatizing words in the language $L = \Sigma^{|x|}$.  This
means that $\Delta u_{\alpha}$ achieves its maximum value
$\frac{k}{\alpha(k+\alpha)}$ as long as $k \leq |x|$.

Given these assumptions, we propose the following procedure for indirectly synthesizing $\mu_{\epsilon,x}$ by sampling a Levenshtin distance $\ell$ and computing the policy $\mu_{\epsilon,x,\ell}$ for only those words that are distance $\ell$ from $x$.

\begin{enumerate}
    \item For a given input word $x$ and an adjacency relation $\adj^s_{w, k}$, 
    fix a desired substitution Levenshtein distance $\ell$ by drawing from the distribution
    \begin{equation} \rho(\ell;|x|,k) = \frac{
    \exp(\frac{\epsilon\alpha(k+\alpha)}{2k(\ell+\alpha)})}{\sum\limits_{\lambda=1}^m \exp(\frac{\epsilon\alpha(k+\alpha)}{2k(\lambda+\alpha)})}
    \end{equation}
    Note that $\rho(\ell;|x|,k) = Pr_{p(w;x)}[d_L(w,x)=\ell]$, the probability of
    selecting an output word $w$ distance $\ell$ from $x$. 
    \item Construct the subset of $\mathcal{A}^{|x|}_{k,x,\ell} \subseteq \mathcal{A}^{|x|}_{k,x}$ that is backwards reachable from $q_{|x|,\ell}$ (the accepting state for words of length $|x|$ and distance exactly $\ell$) and denote the states of $\mathcal{A}^{|x|}_{k,x,\ell}$ as $Q_{k,x,\ell}$
    \item Synthesize $\mu_{\epsilon,x,\ell}: Q_{k,x,\ell} \times \Sigma \to [0,1]$
\end{enumerate}

\begin{algorithm}
\caption{Distance-Restricted Policy Construction}\label{alg::polSyn}
\begin{algorithmic}[1]
\Procedure{SynthesizePolicy}{$\mathcal{A}^{|x|}_{x,k,\ell}$}
\State $V(q_{|x|,\ell}) \gets  1$
\State $CurrQ  \gets  \{q_{|x|,\ell}\}$
\State $counter = 0$
\While{$counter < \ell$}
\For{$q$ s.t. $(q,\sigma,q') \in \delta_{x,\ell}, q' \in CurrQ$}
\State $V(q) \gets \sum_{\{\sigma \in \Sigma| (q,\sigma,q'') \in \delta_{x,\ell}\}} V(q'')$
\State $q \in ActiveQ$
\EndFor
\For {$q \in ActiveQ$, $(q,\sigma,q') \in \delta_{x,\ell}$}
\State $\mu_{x,\ell}(q,\sigma) \gets \frac{V(q')}{\sum_{\{q''|\exists \sigma \in \Sigma (q,\sigma,q'') \in \delta\}} V(q'')}$
\EndFor
\State $CurrQ \gets ActiveQ$
\State $counter \gets counter + 1$
\EndWhile
\Return $\mu_{x,\ell}$
\EndProcedure
\end{algorithmic}
\end{algorithm}

\begin{figure}
\centering
\includegraphics{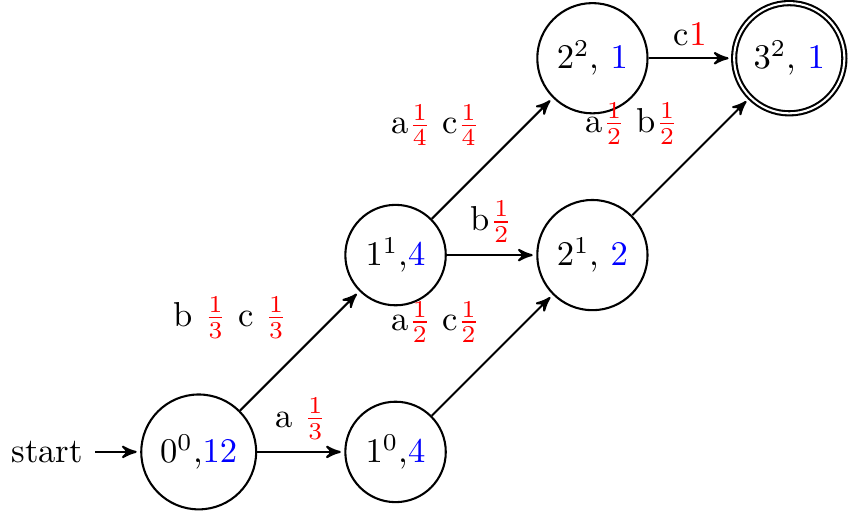}
\caption{Levenshtein automaton for all words of length $3$ and distance  equal to $2$ from the word $abc$ from alphabet $\left\lbrace a, b,c\right\rbrace$.  The value of $V(q)$ is shown in {\color{blue} blue}.  The value of $\mu_{\epsilon,x,\ell}(q,\sigma)$ is shown in {\color{red} red}}
\end{figure}

\subsection{Synthesizing distance-restricted policies}

In this section, we describe the procedure used to construct $\mu_{\epsilon,x,\ell}$.  First, we note that  $p(w;x)$ is a function of $d_L(w,x)$, and thus all strings of the same Levenshtein distance to $x$ should be equiprobable.  In other words,

\begin{equation}
\prod_{e=0}^{|x|-1}\mu_{\epsilon,x,\ell}(q^e,\sigma^e) =\pi_{|x|,\ell,\Sigma} \text{ } \forall q^0\ldots q^{|x|} \in \mathcal{L}(\mathcal{A}_{k,x,\ell})
\end{equation}

 The procedure outlined in Algorithm \ref{alg::polSyn} uses this principle to construct $\mu_{\epsilon,x,\ell}$.  We assign a function $V:Q_{k,x,\ell} \to \mathbb{N}$ such that $V(q)$ is the number of unique paths from $q$ that end in $q_{|x|,\ell}$.  Then, the weighting of the policy at each point is equal to the proportion of unique paths that can be reached by applying the symbol compared to the total number of unique paths reachable from the current state.

\begin{figure}
\centering
\includegraphics{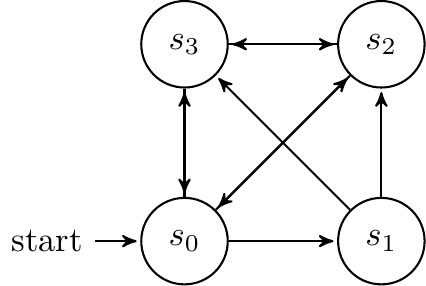}
\caption{\label{fig:TS}Example transition system.  For simplicity, the actions that enable a transition to state $s_i$ is simply labeled $s_i$.  Thus, plans and runs are equivalent for this system.}
\end{figure}

\begin{figure*}
\begin{tabular}{cc}
\includegraphics[scale=0.8]{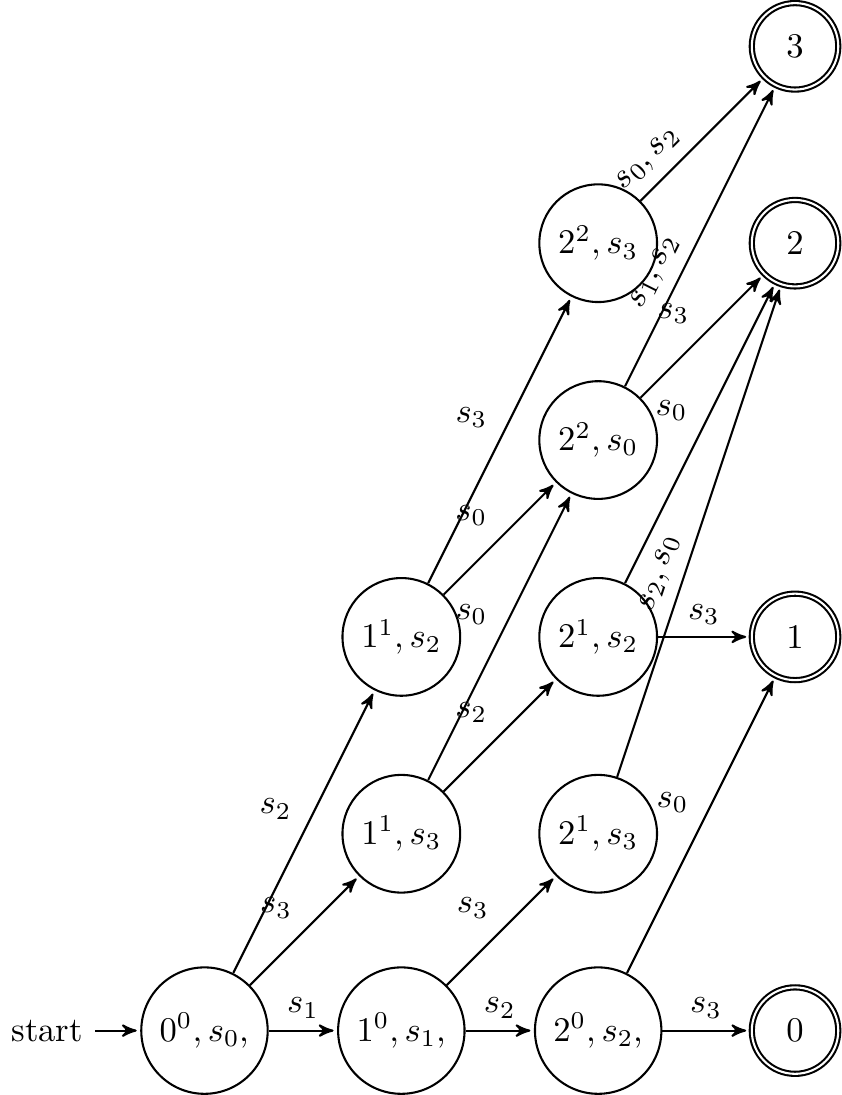} & 
\includegraphics{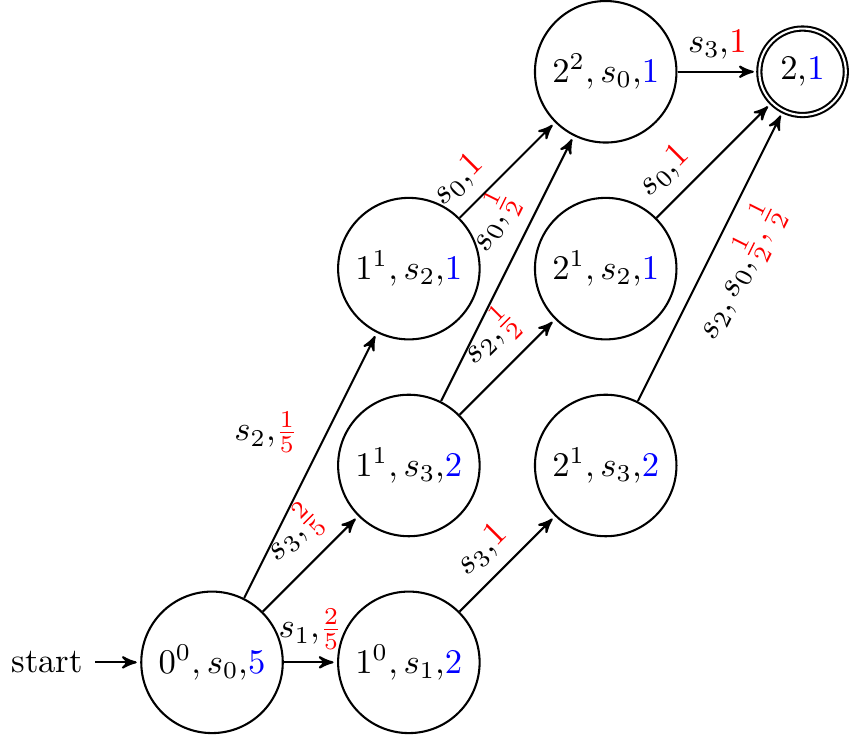}
\end{tabular}
\caption{\label{prodLev}(a) Product Levenshtein automaton for all traces of length $3$ and Levenshtein distance less than or equal to three from  $s_1s_2s_3$  from the transition system shown in Figure \ref{fig:TS}. (b) Restricted product automaton for traces of Levenshtein distance exactly 2 from $s_1s_2s_3$. The value of $V(q)$ is shown in {\color{blue} blue}.  The value of $\mu_{\epsilon,x,\ell}(q,\sigma)$ is shown in {\color{red} red}.} 
\end{figure*}

\section{Generating Differentially Private Runs for a Transition System}
\label{runTS}
Here, we extend the principles used to generate differentially private words presented in Section \ref{wordGen} to generate differentially private runs of a system.  We do this via the introduction of the product Levenshtein automaton.

\begin{dfn}
    Let $\mathcal{A}^{|x|}_{x,k} = (Q,\Sigma,q_0,\delta,F)$ be a Levenshtein automaton and let $TS = (S, s_0, \Sigma, T)$ be a deterministic transition system.  The \emph{Product Levenshtein Automaton}, $\mathcal{A}^{|x|}_{x,k,TS}   = (Q_S,\Sigma,q_0 \times s_0,\delta_{TS},F_{TS})$
    where 
    \begin{itemize}
        \item $Q_S \subseteq Q \times S$
        \item $\delta_{TS} = Q \times S \times \Sigma \times Q \times S$ such that $(q,s,\sigma,q',s') \in \delta_{TS} \Leftrightarrow (q,\sigma,q') \in \delta \wedge T(s,\sigma) = s' $ 
        \item $F_{TS} = \{(q_f,s) \in Q_S | q_f \in F \} $
    \end{itemize}
\end{dfn}

In other words, $\mathcal{A}^{|x|}_{x,k,TS}$ is the \emph{synchronous product} of $\mathcal{A}^{|x|}_{x,k}$ and $TS$.  Further, $w \in \La(\mathcal{A}^{|x|}_{x,k,TS}) \Leftrightarrow  w \in \La(\mathcal{A}^{|x|}_{x,k}) \wedge w \in \La_i(TS)$.  That is, every accepting word in the product corresponds to a sequence of inputs that when applied to $TS$ will result in a run that is within substitution Levenshtein distance $k$ of $w$. An example of a product Levenshtein automaton is shown in Figure \ref{prodLev}(a).   

Because the product Levenshtein automaton is a Levenshtein automaton, we can use the exact same procedure as in Section \ref{wordGen} with using $\mathcal{A}^{|x|}_{x,k,TS}$ instead of $\mathcal{A}^{|x|}_{x,k}$ and ensuring that the maximum distance used to compute $\rho$ is the minimum of $k$ and $\max_{v \in \La_i(TS)} d_L(v,x)$.  An example of applying Algorithm \ref{alg::polSyn} to $\mathcal{A}^{|x|}_{x,k,TS}$ is shown in Figure \ref{prodLev}(b).



\begin{figure*}
\begin{subfigure}[b]{.3\textwidth}
\begin{lstlisting}[linewidth=5.5cm,frame=single]
epsilon=10

amfr ofece91aencctfnot tmf2f2fto
american control conference 2019
american control conference 2019
american control conference 2019
american control conference 2019

epsilon=1

cofrfei0onia2trl11oo21mi20in9er
cler220e 99ntrol aonft0e0a2 o01e
9menic0n cfn0rol n21fereac0 20o9
omr10 c11nnom20ee1ma2 trftci1nt0
0in2mian c0ntroot1onf 01 cn 20l9

epsilon=0.1

c  rrtornil1tmri00ictiro29al2fcc
aoorarailcontroo co1fea9nor 2019
amerioan cocr9rl conference 201i
alerrcan con ofi conierence 2r19
l0t9tn1n ctntroa alffefonce  e12

epsilon=0

9ma9cfc1af2toirom0nt1f2fei29anal
tmerilan conmreliconf reeme na19
cf1 tmnrlcef9rol1mo9fe1l oiim910
fo9 afte 19m9meoenaa0t20roneaf 9
amenmean conlrol ronferenne i01n
\end{lstlisting}
\caption{\label{ACCSamples}  Samples of differentially private versions of the string ``american control conference 2019" generated with different values of the privacy parameter $\epsilon$.} 
\end{subfigure}%
\hspace{0.1\textwidth}
\begin{subfigure}[b]{.6\textwidth}
\begin{tabular}{ccc}
\includegraphics[scale=0.7]{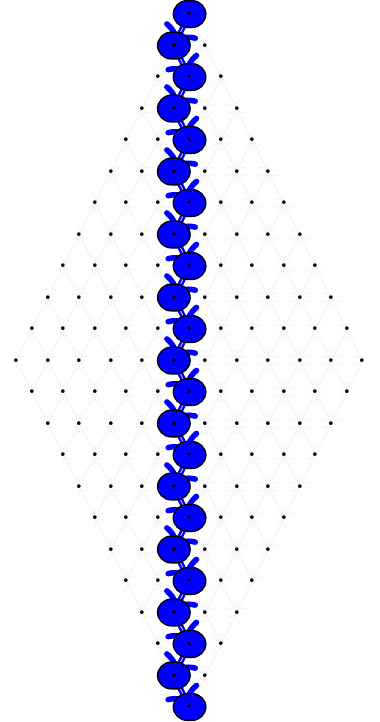} & \includegraphics[scale=0.7]{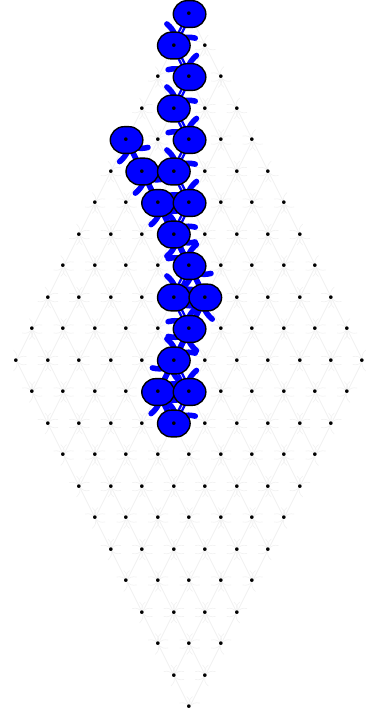} & \includegraphics[scale=0.7]{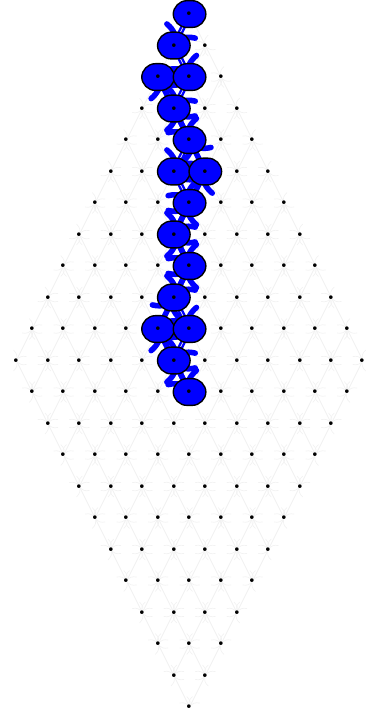} \\
(i) & (ii) & (iii) \\
\includegraphics[scale=0.7]{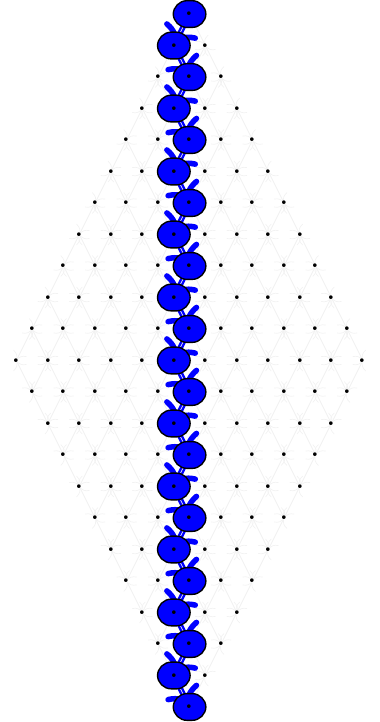} &
\includegraphics[scale=0.7]{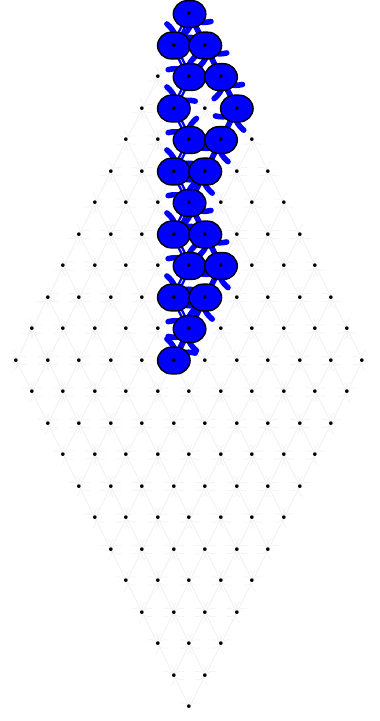} & \includegraphics[scale=0.7]{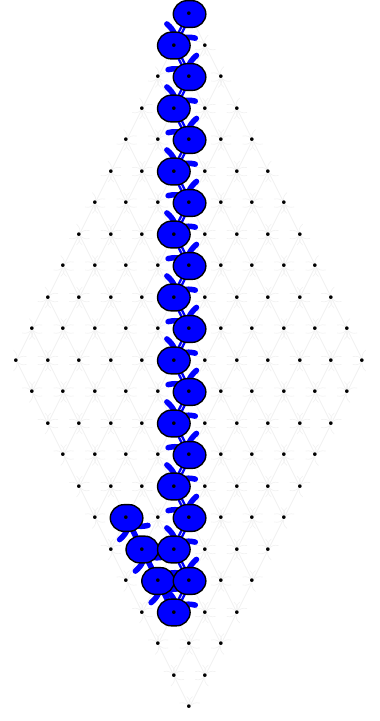} \\
(iv) & (v) & (vi) \\
\end{tabular}
\caption{\label{TsExp} (i) Input trajectory (ii)-(vi) Example trajectories generated by \\ differentially private mechanism with $\epsilon=5$.}
\end{subfigure}
\caption{Numerical results for generate private strings over a finite
alphabet and private runs of a transition system.}
\end{figure*}

\section{Experiments} \label{sec:experiments}
In this section, we present results of some computational experiments
that demonstrate the procedures in Sections \ref{wordGen} and
\ref{runTS}.   We developed a package in the Julia programming
language called LevenshteinPrivacy.jl that implements the procedures
in Sections~\ref{wordGen} and~\ref{runTS}.  
The code uses the LightGraphs.jl framework\footnote{https://github.com/JuliaGraphs/LightGraphs.jl}.  All experiments were performed on a Windows Desktop PC with a 1.90 GHz processor and 16.0GB of RAM.

\subsection{Case Study 1: Strings}
For this set of experiments, we demonstrate the procedure from Section \ref{wordGen} by generating differentially private versions of the string ``american control conference 2019".  The alphabet is comprised of all the unique characters in the input string.  The Levenshtein automaton c contains 561 states and 1056 edges.  The automaton was constructed in 6.6s and generating 40 privatized strings required 1.53s of computation time.  Figure \ref{ACCSamples} shows outputs from these experiments with different values of the privacy parameter $\epsilon$. As we can see, as $\epsilon$ decreases 
(strength of privacy increases), the outputs become less recognizable until they become almost entirely gibberish.

\subsection{Case Study 2: Transition System}

 In this case study, we demonstrate the procedure outlined in Section \ref{runTS}.  We constructed a transition system with 225 states and edges that corresponds to a 15 by 15 ``grid world".  The path we wish  to privatize is shown in Figure \ref{TsExp}(i). The resulting product automaton constructed by Algorithm has 6194 states and 18762 edges.   The computation time for constructing the product automaton was 193s and the time required to generate 100 samples with $\epsilon=0.01$ was 6.4s.

\section{Conclusions and Future Work} \label{sec:conclusion}
In this work we presented a method for providing differential privacy
to words over a finite alphabet. An exponential mechanism was
devised to generate possible output words, and the theory of
Levenshtein automata was applied to efficiently generate samples
from this distribution. Numerical results validated these theoretical
developments and demonstrated their efficiency.

The first natural extension of this work is to the full Levenshtein distance,
which will allow for not only substitutions as in this work, but also deletions
and insertions. A key challenge in doing so is efficiently generating
samples from the distribution over possible outputs. This work did so
by considering strings of a fixed output length, which corresponds
to using the substitution Levenshtein distance rather than the full
Levenshtein distance. Making this extension will require fundamental
innovations beyond this work, though successfully making this
extension will significantly broaden the scope of this work. 

\bibliographystyle{unsrt}
\bibliography{mybib}

\begin{thebibliography}{10}

\bibitem{simmhan11}
Y.~Simmhan, A.~G. Kumbhare, B.~Cao, and V.~Prasanna.
\newblock An analysis of security and privacy issues in smart grid software
  architectures on clouds.
\newblock In {\em 2011 IEEE 4th International Conference on Cloud Computing},
  pages 582--589, July 2011.

\bibitem{siddiqui12}
F.~Siddiqui, S.~Zeadally, C.~Alcaraz, and S.~Galvao.
\newblock Smart grid privacy: Issues and solutions.
\newblock In {\em 21st Int. Conf. on Computer Communications and Networks
  (ICCCN)}, pages 1--5, July 2012.

\bibitem{cortes16}
J.~Cort{\'e}s, G.~Dullerud, S.~Han, J.~Le~Ny, S.~Mitra, and G.~Pappas.
\newblock Differential privacy in control and network systems.
\newblock In {\em Decision and Control (CDC), 2016 IEEE 55th Conference on},
  pages 4252--4272, 2016.

\bibitem{wang17}
Yu~Wang, Zhenqi Huang, Sayan Mitra, and Geir~E Dullerud.
\newblock Differential privacy in linear distributed control systems: Entropy
  minimizing mechanisms and performance tradeoffs.
\newblock {\em IEEE Transactions on Control of Network Systems}, 4(1):118--130,
  2017.

\bibitem{nozari18}
E.~Nozari, P.~Tallapragada, and J.~Cort{\'e}s.
\newblock Differentially private distributed convex optimization via functional
  perturbation.
\newblock {\em IEEE Trans. on Control of Network Systems}, 5(1):395--408, 2018.

\bibitem{hale17}
Matthew~T Hale and Magnus Egerstedt.
\newblock Cloud-enabled differentially private multi-agent optimization with
  constraints.
\newblock {\em IEEE Transactions on Control of Network Systems}, 2017.
\newblock In press.

\bibitem{han17}
Shuo Han, Ufuk Topcu, and George~J Pappas.
\newblock Differentially private distributed constrained optimization.
\newblock {\em IEEE Transactions on Automatic Control}, 62(1):50--64, 2017.

\bibitem{hale18}
M.~Hale, A.~Jones, and K.~Leahy.
\newblock Privacy in feedback: The differentially private lqg.
\newblock In {\em 2018 Annual American Control Conference (ACC)}, pages
  3386--3391, June 2018.

\bibitem{leny14}
J.~Le Ny and G.~J. Pappas.
\newblock Differentially private filtering.
\newblock {\em IEEE Transactions on Automatic Control}, 59(2):341--354, Feb
  2014.

\bibitem{leny18}
J.~Le Ny and M.~Mohammady.
\newblock Differentially private mimo filtering for event streams.
\newblock {\em IEEE Transactions on Automatic Control}, 63(1):145--157, Jan
  2018.

\bibitem{tabuada09}
Paulo Tabuada.
\newblock {\em Verification and control of hybrid systems: a symbolic
  approach}.
\newblock Springer Science \& Business Media, 2009.

\bibitem{belta07}
C.~Belta, A.~Bicchi, M.~Egerstedt, E.~Frazzoli, E.~Klavins, and G.~J. Pappas.
\newblock Symbolic planning and control of robot motion.
\newblock {\em IEEE Robotics Automation Magazine}, 14(1):61--70, March 2007.

\bibitem{dwork14}
Cynthia Dwork, Aaron Roth, et~al.
\newblock The algorithmic foundations of differential privacy.
\newblock {\em Foundations and Trends{\textregistered} in Theoretical Computer
  Science}, 9(3--4):211--407, 2014.

\bibitem{dwork06}
C.~Dwork, K.~Kenthapadi, F.~McSherry, I.~Mironov, and M.~Naor.
\newblock Our data, ourselves: Privacy via distributed noise generation.
\newblock In {\em Annual International Conference on the Theory and
  Applications of Cryptographic Techniques}, pages 486--503. Springer, 2006.

\bibitem{mcsherry07}
F.~McSherry and K.~Talwar.
\newblock Mechanism design via differential privacy.
\newblock In {\em 48th Annual IEEE Symposium on Foundations of Computer Science
  (FOCS'07)}, pages 94--103, Oct 2007.

\bibitem{eigner14}
Fabienne Eigner, Aniket Kate, Matteo Maffei, Francesca Pampaloni, and Ivan
  Pryvalov.
\newblock Differentially private data aggregation with optimal utility.
\newblock In {\em Proceedings of the 30th Annual Computer Security Applications
  Conference}, pages 316--325. ACM, 2014.

\bibitem{hardt12}
Moritz Hardt, Katrina Ligett, and Frank McSherry.
\newblock A simple and practical algorithm for differentially private data
  release.
\newblock In {\em Advances in Neural Information Processing Systems}, pages
  2339--2347, 2012.

\bibitem{hay09}
M.~Hay, C.~Li, G.~Miklau, and D.~Jensen.
\newblock Accurate estimation of the degree distribution of private networks.
\newblock In {\em 2009 Ninth IEEE International Conference on Data Mining},
  pages 169--178, Dec 2009.

\bibitem{kasiviswanathan13}
Shiva~Prasad Kasiviswanathan, Kobbi Nissim, Sofya Raskhodnikova, and Adam
  Smith.
\newblock Analyzing graphs with node differential privacy.
\newblock In Amit Sahai, editor, {\em Theory of Cryptography}, pages 457--476,
  Berlin, Heidelberg, 2013. Springer Berlin Heidelberg.

\bibitem{pinot18}
Rafael Pinot, Anne Morvan, Florian Yger, C{\'e}dric Gouy-Pailler, and Jamal
  Atif.
\newblock Graph-based clustering under differential privacy.
\newblock {\em arXiv preprint arXiv:1803.03831}, 2018.

\bibitem{levenshtein}
Vladimir~I Levenshtein.
\newblock Binary codes capable of correcting deletions, insertion, and
  reversals.
\newblock {\em Soviet physics doklady}, 10(8):707--710, 1966.

\bibitem{schulz}
Klaus~U Schulz and Stoyan Mihov.
\newblock Fast string correction with levenshtein automata.
\newblock {\em International Journal on Document Analysis and Recognition},
  5(1):67--85, 2002.

\bibitem{hamming50}
R.~W. Hamming.
\newblock Error detecting and error correcting codes.
\newblock {\em The Bell System Technical Journal}, 29(2):147--160, April 1950.

\bibitem{wagner74}
Robert~A. Wagner and Michael~J. Fischer.
\newblock The string-to-string correction problem.
\newblock {\em J. ACM}, 21(1):168--173, January 1974.

\end{thebibliography}

\end{document}